# Equation of State of Colloidal Membranes

Andrew J. Balchunas[a], Rafael A. Cabanas[a], Mark J. Zakhary[a], Thomas Gibaud[b], Seth Fraden[a], Prerna Sharma[c], Michael. F. Hagan[a] and Zvonimir Dogic[a,d]

a. Department of Physics, Brandeis University, Waltham, MA 02454
b. Unversité de Lyon, Ens de Lyon, Université Claude Bernard, CNRS, Laboratoire de Physique, F-69342 Lyon
c. Department of Physics, Indian Institute of Science, Bangalore 560012, India
d. Department of Physics, University of California at Santa Barbara, Santa Barbara, CA 93106

In the presence of a non-adsorbing poylmer, monodisperse rod-like colloids assemble into one-rod-length thick liquid-like monolayers, called colloidal membranes. The density of the rods within a colloidal membrane is determined by a balance between the osmotic pressure exerted by the enveloping polymer suspension and the repulsion between the colloidal rods. We developed a microfluidic device for continuously observing an isolated membrane while dynamically controlling the osmotic pressure of the polymer suspension. Using this technology we measured the membrane rod density over a range of osmotic pressues than is wider that what is accesible in equilibrium samples. With increasing density we observed a first-order phase transition, in which the in-plane membrane order transforms from a 2D fluid into a 2D solid. In the limit of low osmotic pressures, we measured the rate at which individual rods evaporate from the membrane. The developed microfluidic technique could have wide applicabilty for *in situ* investigation of various soft materials and how their properties depend on the solvent composition

## Introduction

Colloidal membranes are one-rod-length thick monolayers of aligned rods that assemble in a presence of depleting polymer [2-5]. They represent a robust pathway for assembly of self-limited structures that does not rely on the chemical heterogeneity of the amphiphilic building blocks, but rather on their geometry. Numerous properties of colloidal membranes, including their out-of-plane bending rigidity, are determined by the in-plane density of the rodlike constituents[2,6]. In turn, this density is determined by the osmotic pressure exerted by the enveloping depletant that is balanced by the electrostatic repulsions of the charged rod-like viruses. We describe a microfluidic platform that allows us to continuously change the depletant concentration, while simultaneously measuring the membrane area using optical microscopy. From such data we reconstruct the colloidal membrane equation of state, which relates the osmotic pressure exerted on the membrane to the effective in-plane rod density. With increasing osmotic pressure the equation of state exhibits a discontinuity due to a 2D liquid-to-solid phase transition of the constituent rods.

Colloidal membranes with in-plane crystalline order were studied previously. Early studies have demonstrated that in the presence of small molecular weight depletant, filamentous viruses assembled into hexagonally shaped membranes[7]. More recently, X-ray scattering analysis confirmed crystalline order for these conditions [8]. Solid membrane have also been observed in a mixture of virus and poly-ethylene-glycol, whose osmotic pressure is temperature dependent [9]. This feature made it possible to induce nucleation and crystallization *in situ*, which was accompanied by large 3D out-of-plane membrane deformations. Another work examined the liquid to crystal transition with X-ray scattering techniques [10]. Our measurements reveal the magnitude of the discontinuous volume change at the transition, and allow us to fit the entire equation of state to a theoretical model. These results have far-reaching implications for understanding colloidal membranes. For example, the out-of-plane bending modulus of a colloidal membrane is characterized by the local compression or expansion of molecules, which increases away from the stress-free neutral surface[11]. Thus, the out-of-plane membrane deformations are intimately coupled to its in-plane compressibility modulus. The equation of state yields the lateral compressibility, which in turn provides an estimate of the curvature modulus of the colloidal membranes.

**Materials and Methods:**

**Microfluidics device for *in-situ* buffer exchange**: Colloidal membranes are fragile structures held together by weak osmotic pressures. Even the slightest flows, or the movement of an AFM tip close to the surface can fragment a membrane. To overcome these challenges we designed a microfluidic device that exchanges the buffer with minimal flow distortions, while allowing for continuous observation of the assemblage with high numerical aperture optics. Technologies have been developed that equip microfluidic devices with dialysis membranes that enable rapid spatial and temporal buffer exchange[12, 13]. However, our experiments impose an additional constraint because of the need to exchange large molecular weight depletant polymers that do not pass through conventional dialysis membranes.

Our microfluidic device had two PDMS layers bound together. The first layer contained the long channel with stepped notches on both sides, while a second layer contained pressurised channels that act as valves that open and close the flow channel (**Fig. 1**) [14, 15]. The notches reduced the laminar flow velocity. Due to the continuity of the flow streams, larger channel cross-sections reduced flow velocities. We performed experiments in notches with one or two steps, but membranes stored in a notch with no steps were typically destroyed by the laminar flow through the main channel. A critical element was the inclusion of the second layer that contained two on-chip valves. When closed, these effectively blocked minute flows during the many-hour long interval that is necessary for the membranes to grow to a sufficient diameter. At a flow rate of 10 μL per hour, a complete buffer exchange took approximately ten minutes, including the time for the depletant to diffuse from the main channel into the notches (**Fig. 2**).

The experiment required maintaining the same buffer composition over many hours. To accomplish this, the device was designed with a cavity that encircled the channel and notches and acted as a aqueous reservoir that is included in the first layer. This PDMS layer was sandwiched between a microscope glass slide and a plastic COC slab with low water permeability, thereby minimizing the evaporation of water from the sample chamber to the outside (**Fig. 1**). A hydrostatic pump was created by closing the outlet and connecting the inlet to a solvent filled tube that was positioned 1 m above the sample. This replenished the evaporated solvent due to PDMS permeability. The main channel that contained the sample and the water reservoir were not directly connected. The water reservoir was critical because of the long duration of the experiments and the small sample volumes. The samples were visualized with high numerical aperture objective. The device thickness required a long working distance condenser for brightfield and differential interference contrast (DIC) microscopy. High resolution images of the membranes were obtained using fluorescence.

Directly flowing fluid into the chamber resulted in air bubbles trapped in the notches. Eliminating air from the device was accomplished through dead-end filling the device by closing the outlet valve and then using a syringe pump to load the chamber. As PDMS is highly permeable to air, the chamber was evacuated by the air diffusing through the closed valves. Once loaded with a rod-polymer mixture, the formation of membranes with diameter of tens of microns required about 24 hours.

Glass coverslips (GoldSeal, FisherScientific) were coated with the acrylamide polymer brush to prevent virus adsorption[16]. Plasma treatment usually used for bonding PDMS to glass destroyed the polymer-brush. We tried, but failed, to grow the polymer brush after bonding the PDMS to glass. Therefore, we clamped together the PDMS chip to the acrylamide treated glass coverslip using an aluminium frame (**Fig. 1a**). Every component in the device can be washed and reused. To avoid bending deformations of the thin PDMS layers when clamping the device, a slab of rigid Mylar plastic and a shaped shim plate of COC plastic layer were used to apply pressure directly over the chip elements forming the seal. To ensure uniform compression of the layers, and to provide compliance while sealing, a soft rubber O-ring was placed on top of the stack in contact with the aluminum clamp. The aluminum clamp was tightened using four screws. Enough pressure had to be applied to avoid leakages from the main channel or the reservoir, but too much pressure yielded collapsed channels.

Manufacture of the microfluidic device required two photoresist masters that were used to mold the PDMS device. The masters were fabricated using conventional soft photolithography techniques[17]. The flow channel and the reservoir were produced from one master, with photoresist features of two different heights with one using positive photoresist and the other negative photoresist. The inlet and the outlet valves were deposited in a positive photoresist to achieve a ~10 μm height. The main channel with the notches and the reservoir, were deposited using negative photoresist with ~50 μm height. A separate wafer encoded the control layer for the two valves. Once molded into the PDMS, these pressure driven control valves were positioned above the inlet and outlet valves. The PDMS device was fabricated by spin coating a thin layer of PDMS over the wafer that contained the inlet, outlet, main channel and reservoir. A several mm thick layer of PDMS was poured over the wafer that contains the features of the valves. Unlike the original Quake formulation[14], we used 1:10 ratio of PDMS components for both layers. Once cured, both layers were bonded using plasma, and inlet and outlet fill-holes were punched into the PDMS. This way, the inlet, outlet, main channel with notches and reservoir were exposed and clamped directly on top of the microscope slide, while the valves that open and close the inlet and outlet channels where situated directly above and oriented perpendicular across those channels. (Fig. 1)

**Optical microscopy, data acquisition and analysis:** Experiments were performed using an inverted optical microscope (Nikon Eclipse Ti) equipped with a long working distance (LWD) condenser, and a 100x oil immersion objective (PlanFluor, NA 1.3). A optical tweezer, created by focusing a 1064 nm laser (Compass 1064, Coherent), and steered with a pair of acoustic-optic modulators was used to position membranes in the microfluidic notches[18]. Images were acquired using a cooled CCD camera (Clara, Andor) controlled by Micro-Manager. The colloidal membranes were visualized either with DIC or fluorescence microscopy. For fluorescence images, the microscope was equipped with a mercury light source (X-Cite 120Q) and a single band dichroic filter (LED-TRITC-A-NTE-ZERO, SemRock).

**Sample Preparation:** Bacteriophage *fd*-wt (wild-type) and *fd*-Y21M were purified using standard methods[19]. The viruses were prepared at isotropic-nematic phase coexistence, and the experiments were performed by only using viruses from the isotropic fraction. Shorter rods preferentially partition into the isotropic phase, hence this procedure reduced the number of long, end-to-end fused oligomers that destabilize colloidal membranes[2]. A depleting agent, Dextran MW 500,000 (Sigma-Aldrich, D5251, Lot #023K0675), was mixed with viruses at conditions known to promote assembly of colloidal membranes. All samples were made in 100 mM NaCl, 20 mM Tris-HCl, pH 8.0 buffer, and had a final virus concentration of 2.5 mg mL$^{-1}$. Due to the difference in persistence length of *fd*-wt and *fd*-Y21M, the initial amount of depletant was different for each virus type [20]: *fd*-wt and *fd*-Y21M membranes were assembled at depletant concentrations of 42 mg mL$^{-1}$ and 37 mg mL$^{-1}$, respectively. The mixture was transferred into tubing (PTFE #30 AWG, Cole-Parmer) using a syringe (5 µL, Hamilton Gastight), and a syringe pump (PHD 22/2000, Harvard Apparatus) was used to dead-end fill the PDMS device. The sample was always introduced at constant flow-rate, so the pressure in the device channels is unknown. Fluorescent membranes were assembled with viruses labelled with an amine reactive fluorescent dye (DyLight-550 NHS Ester, FisherScientific)[21]. There were about 20 dye molecules per virus as determined by the relative intensity of the absorbance at the wavelength related to the virus (269 nm) and the fluorescent dye (550 nm). Samples used to characterize the dynamics of the constituent rods were doped with labelled viruses at a ratio of 1 to 50,000 of labelled to unlabelled particles. Fluorescence images revealed the dynamics of individual rods within the membrane.

**Image Analysis:** The membrane area was measured using two methods. First, we determined the area of a fluorescently labelled membrane with ImageJ, by setting an intensity threshold at two standard deviations above the mean pixel intensity. Second, we used DIC microscopy to visualize the membrane's edge, and from there we measured the membrane area with the circle tool in ImageJ. Both measurements were consistent with each other. Due to its monolayer structure, the membrane area is directly proportional to its volume. To convert the measured area into density, we analysed a few equilibrium membrane samples with small angle X-Ray (SAXS) measurements, and extracted the particle separation. Interactive data language (IDL, ITT VIS) software was used to track individually labelled viruses, and from there measure the mean square displacement for isolated rods diffusing within a membrane.

**SAXS Experiments:** Small angle X-ray scattering (SAXS) experiments on individual membranes were performed at the ID02 beamline of the European Synchrotron Radiation Facility in Grenoble, France [22]. The samples were held in the cover slip chamber with membranes lying flat on the coverslip. The chamber was set perpendicular to the X-ray device, so that the membranes were in a face-on configuration with respect to the incoming beam. The sample was scanned with a 20 x 20 µm$^2$ X-ray beam for individual colloidal membranes, and signal was recorded with a *q* range between 0.04 nm$^{-1}$ to 2 nm$^{-1}$. The intensity measured in the supernatant was used as the background signal. The plotted intensities were radial averages of the difference between the scattered intensity from the membrane and the background intensity.

X-Ray experiments yield a scattered intensity *I(q)* with a correlation peak positioned at $q_{peak}$, which is related to average lateral filament separation (Fig. 5b). To estimate the average distance between the filaments we need to divide the measured intensity by the form factor *F(q)* which would yield the structure factor: S(q)~I(q)/F(q). The form factor for our experiment would consist of a dilute suspension of rods that are all aligned along the along the direction of the incident x-ray beam. In practice, this is not doable as dilute rod suspensions are necessarily isotropic (Fig. 5a). To make progress we theoretically estimate the form factor and from there calculate the structure factor of colloidal membranes. Note that the division by the form factor shifts the peak location by at most 3% (Fig. 5b). From the measured structure factor we estimate the average filament separation $d = 2\pi/q_{peak}$. Assuming local hexagonal packing yields the density of rods within *fd*-wt membranes.

To normalize the *fd*-Y21M equation of state, we estimated the densities of these membranes using DIC microscopy. The DIC signal along the shear axis is proportional to the index of refraction difference between the membrane and the buffer. In turn, the membrane index of refraction is linearly proportional to the virus density. The edge intensity along the shear axis and the membrane's rod density should be linearly related. Comparing this data with the DIC edge intensity of a *fd-wt* membrane of a known density, yielded the absolute density of *fd*-Y21M membranes. Repeating this analysis on *fd-wt* membranes with known densities from X-ray diffraction measurements revealed that this method has ~10% accuracy.

**Experimental Results:**

After mixing the virus and polymer suspension, one-rod length thick mesoscopic disk-like structures nucleated in the bulk suspension. Within a few minutes, viruses completely phase separated from the isotropic polymer suspension. Thereafter, the mesoscale disks grew into mature membranes through lateral coalescence[23]. After reaching a certain size, the dense membranes sedimented on to the coverslide, ensuring that they remained in the image plane for the duration of the experiment (**Fig. 3**). We formed colloidal membranes within a microfluidic device. Subsequently, using an optical tweezer, we placed a few isolated circular membranes within a stepped notch on a side of the main channel. The membranes were continuously visualized with either fluorescence, brightfield, or DIC microscopy. For conditions where equilibrium membranes formed, the mature membranes had a constant area as long as the buffer composition remained the same, indicating that the evaporation of the virus into the isotropic suspension was negligible on the experimental time scales. Consequently, any change in the membrane area in response to changing osmotic stress, or equivalently depletant concentration, was directly related to the change in the virus density, within an unknown scaling factor.

We performed a sequence of buffer exchanges. After each exchange, we waited ~60 minutes to allow the depletant to diffuse into the stepped notch and the sample to equilibrate, before measuring the membrane area. The area significantly changed with depletant concentration. For example, increasing the Dextran concentration from 33.5 mg/ml to 100 mg/ml reduced the area by 42% (**Fig. 5**). We measured the area of the same membrane for ~20 different dextran compositions. In order to average data from different membranes all measurement sweeps had at least one data point with a common Dextran concentration, which was used to rescaling the area.

The above-described procedure yielded the membrane area as a function of the applied osmotic pressure. Converting this data into an equation of state requires the scaling factor that relates area to virus density. Therefore, we have measured the virus density within a membrane by using small angle X-Ray scattering, which yielded an isotropic ring whose peak location can be used to calculate the average lateral filament separation (**Fig. 4**). The scattering experiments were repeated at a few dextran concentrations for which equilibrium membranes formed, and the densities found in the x-ray were consistent with those found in the microfluidic measurements.

Equilibrium colloidal membranes of *fd-wt* formed over a fairly narrow range of dextran concentrations (from 40 mg/ml to 55 mg/ml). Microfluidic device allowed for measurements over a significantly larger range of osmotic pressures (**Fig. 6a**). This was possible because 2D colloidal membranes are highly metastable structures that cannot readily melt into 3D nematic tactoids at lower depletant concentrations, or freeze into 3D smectic liquid crystals at high concentrations. The equation of state exhibited a pronounced discontinuity at 55 mg/ml dextran concentrations, where the area changed by 8%. This is suggestive of a first order phase transition, due to two-dimensional freezing of the in-plane virus order. At osmotic pressures below the transition, colloidal membranes exhibited significant edge fluctuations due to finite edge tension[24]. Increasing the dextran concentration suppressed these fluctuations, and above the transition they entirely ceased. The arrest of edge fluctuations was accompanied by a discontinuous change in the membrane area. Doping the membrane with fluorescent rods revealed the dynamics of constituent rods in both phases. At lower osmotic pressures viruses exhibited diffusive motion, but above the discontinuity their motion ceased (**Fig. 7**). This confirms our hypothesis that with increasing osmotic pressure colloidal membranes undergo a freezing transition from a liquid-like to solid-like membranes. The transition is reversible, with no measurable hysteresis.

We also measured the equation of state experiment for *fd*-Y21M membranes, a virus with a point mutation in the major coat protein (**Fig. 6b**). This subtle structural change yields filaments with essentially the same contour length, but a larger persistence length of 9.9 $\mu$m [20,25]. Both *fd*-wt and *fd*-Y21M have similar equations of state, although the dextran concentration required to induce a phase transition into a 2D crystal is different for the two virus types. Fd-wt membranes crystallized at ~55 mg/ml while *fd*-Y21M crystalized at ~42.5 mg/ml. We ascribe the difference in the transition pressure to the difference in persistence length of the viruses.

We investigated the upper and lower bounds over which the colloidal membranes remain metastable. At sufficiently high osmotic pressures we found that colloidal membranes solidified and fractured. Notably the onset of fracture depended on virus type: *fd-wt* fractured at much lower concentrations compared to *fd*-Y21M. At lower dextran concentrations we found two different pathways by which membranes became unstable. In one pathway, we observed direct nucleation of 3D nematic tactoids. The nucleation event induced formation of large-scale flows that quickly transported all the rods from a metastable 2D membrane into a more stable 3D tactoid (**Fig. 8**). Intriguingly the tactoid nucleation events only occurred at the membrane's edge, where the boundary conditions induce tilt and the associated formation of a thin edge-bound layer of a quasi 1D-nematic phase [23,24]. The alternate pathway is direct evaporation of rods into the background isotropic suspension at low enough osmotic pressures (**Fig. 9**).

To investigate rod evaporation, we formed *fd*-Y21M membranes in the stable regime at a dextran concentration of 40 mg/ml. Subsequently, we performed a buffer exchange to a dextran concentration to 25 mg mL$^{-1}$, a regime where the equation of state could not be measured since the membrane area changes on experimental time scales. The buffer exchange took ~10 minutes, while the membrane area decreased on a longer time scale of tens of minutes (**Fig. 9**). In principle, virus particles can evaporate either from the edge bound 1D nematic, or directly from the 2D membrane interior. The former case would yield the area rate of change proportional to the membrane circumference, or equivalently the square root of the area. For the latter case, the evaporation rate would be proportional to the area. The experimentally measured curve was well described by the exponential curve, which indicates that virus mainly evaporated throughout the entire membrane interior. The unbinding rate for the viruses under these conditions is on average 4.1 min$^{-1}$.

**Theoretical model of the equation of state:** We develop a phenomenological model that explains some aspects of the measured equation of state. We assume that there is no depletant inside the membrane. Therefore, mechanical equilibrium requires that the applied osmotic pressure, $\Pi_{\text{app}}$, be balanced by the pressure of the rods within colloidal membrane, $\Pi_{\text{int}}$. We estimate the osmotic pressure arising from inter-rod electrostatic interactions and the suppression of rod conformational fluctuations as the rod density increases using a previous theoretical model[26]. In brief, we calculate the lateral extent of conformational fluctuations, *d*, of a filament within a confining electrostatic field generated by nearest neighbour rods arranged on a hexagonal array. The characteristic length along the rod contour of such undulations is given by the deflection length:

$$l_{\text{d}} \cong d^{2/3} l_{\text{p}}^{1/3} \qquad (1)$$

with $l_{\text{p}}$ the filament persistence length[27-29]. Each deflection gives rise to a free energy cost that scales with $k_B T$, so the free energy per rod arising from conformational fluctuations is proportional to the number of deflections over the total filament contour length:

$$F_{\text{conf}} = c k_{\text{B}} T / l_{\text{d}} \qquad (2)$$

with the constant given by $c = 2^{-2/3}$ [26].

We assume that the deflection length is large compared to the Debye length $l_{\text{d}} \gg \kappa^{-1}$ so that the filaments can be locally treated as rigid rods when calculating the electrostatics. In this limit, the far field of the electrostatic potential $\phi$ from a rod (outside of its double layer) is described by the Debye-Hückel approximation, which for a cylinder is given by:

$$\psi(\kappa r) \sim 2\, \xi_{\text{eff}} K_0(\kappa r), \text{ for } r \geq \frac{\sigma}{2} + \kappa^{-1}, \qquad (3)$$

where $\psi = e\phi/k_B T$ is the non-dimensional electrostatic potential, *e* is the elementary charge, $K_0$ is the zero-order modified Bessel function of the second kind, and $\xi_{\text{eff}} = l_{\text{B}} \nu_{\text{eff}}$ is the dimensionless linear charge density, with $l_{\text{B}} \approx 0.71$ nm the Bjerrum length and $\nu_{\text{eff}}$ the effective filament linear charge density [30]. The effective charge density can be calculated from the nonlinear Poisson-Boltzmann equation as described below.

The electrostatic free energy depends sensitively on the scale of lateral undulations *d*, since these bring neighbouring rods closer together, thus enhancing the electrostatic energy. Performing a variational calculation, in which *d* is determined by minimizing the total free energy $F_{\text{tot}}$ arising from conformational fluctuations and electrostatics, results in the following expression for *d* [26]:

$$\frac{d^{8/3}e^{\frac{1}{2}\kappa^2 d^2}}{1+\frac{1}{2}\kappa^2 d^2 d_{\text{rod}}^{-1}} = \frac{2cl_B d_{\text{rod}}^{1/2} e^{\kappa d_{\text{rod}}}}{9(2\pi)^{1/2}\xi_{\text{eff}}^2 l_p^{1/3}\kappa^{3/2}}. \tag{4}$$

Assuming hexagonal ordering of the rods, the osmotic pressure is given by: $\Pi_{\text{int}} = -\frac{\partial F_{\text{tot}}}{3^{1/2}d_{\text{rod}}\partial d_{\text{rod}}}$ [31], which yields:

$$\Pi_{\text{int}} = \frac{2ck_B T}{3^{2/3}\kappa d_{\text{rod}} d^{8/3} l_p^{1/3}}. \tag{5}$$

A subsequent work extended the Odijk's calculation and relaxed some approximations; however, we were not able to match the scaling of the experimental applied osmotic pressure using this model[32]. Moreover, an earlier calculation within a hexagonal layer of charged filaments predicts that the osmotic pressure *increases* with persistence length[33], in contrast to our experiments which exhibit a decrease in pressure with persistence length. This calculation is inappropriate for the semiflexible regime because it assumes a persistence length that is small or on the order of the inter-rod spacing, as has been previously noted[26]. A more recent calculation for the effect of fluctuations on inter-rod interactions obtains a decreasing dependence of osmotic pressure with persistence length[35]. However, we do not observe the predicted doubling or quadrupling of the apparent decay length of the interactions due to fluctuations for the range of inter-rod separations studied in our experiments, although this could become relevant for larger inter-rod separations.

To proceed further, we need to solve for the effective linear charge density $\xi_{\text{eff}}$. Due to the finite width of the *fd*-virus, the estimate for an infinitely thin cylinder that counterions will renormalize the charge density to one charge per Bjerrum length is inapplicable[36]. Instead, we note that the Debye-Hückel approximation accurately describes the far-field form of the electrostatic potential, but over-predicts the potential in the near field[28, 37]. Therefore, we find the effective charge density for which the far-field potential is correct, as has been described previously. We used an approximate analytical solution to the nonlinear Poisson-Boltzmann equation around a cylinder, which matches a near-field solution to the Debye Hückel far-field[38]. Equating the far-field result to Eq. 3 yields the effective charge density, as a function of the bare charge density $v_0$ and $\kappa$.

To capture the experimental observation that the decay length of the interactions is shorter than the bulk Debye length, it was necessary to account for the presence of excess counterions within the colloidal membrane. We related the local Debye length $\kappa^{-1}$ to the bulk value $\kappa_D^{-1}$ and the membrane charge density using the cylindrical cell model[32,39]. In this approach, the hexagonal Wigner–Seitz cell associated with each virus is represented by a cylinder with the same volume. The radius of the effective cylindrical Wigner–Seitz cell is:

$$R_S = d_{\text{rod}}\sqrt{\sqrt{3}/2\pi}, \tag{6}$$

and the local Debye length is calculated as described previously[32],

$$\kappa = \kappa_D\sqrt{\cosh\psi_S}, \tag{7}$$

and

$$\tanh\psi_S = \frac{2\xi_{\text{eff}}\big(I_0(\kappa R_S)K_1(\kappa R_S)+I_1(\kappa R_S)K_0(\kappa R_S)\big)}{I_1(\kappa R_S)-K_1(\kappa R_S)\frac{I_1(\kappa\sigma/2)}{K_1(\kappa\sigma/2)}} \tag{8}$$

where $\psi_S$ is the dimensionless electrostatic potential at the surface of the cell. Since $\xi_{\text{eff}}$ and $\kappa$ are interrelated, it is necessary to solve for them self-consistently as functions of $d_{\text{rod}}$.

We set the *fd* diameter as $\sigma = 6.6$ nm, and use the estimate of $v_0 = -7$ e / nm for *fd*[40]. For the bulk Debye length $\kappa_D^{-1} = 1$ nm, we obtain an effective dimensionless linear charge density of $\xi_{\text{eff}} = 36.15$. For comparison, using the Debye Huckel approximation without the near-field correction would result in an effective charge density of $\xi_{\text{DH}} = \frac{l_B}{b\kappa(\sigma/2)K_1(\kappa\sigma/2)} \approx 53.56$. The Debye screening length and effective filament charge depend on the filament separation (**Fig. 10**). The local Debye length varies by 40% over the experimental range of inter-rod distances.

To compare the experimental data to the theoretical model we converted the dextran concentration into the osmotic pressure using a modification of the previously published empirical relationship:

$$\Pi(c) = A_1(m_{w0}/m_w)c + A_2c^2 + A_3c^3,$$

where $c$ is the dextran weight fraction, $A_1$=0.0655 atm cm³ gm⁻¹, $A_2$=10.38 atm cm⁶ gm⁻², $A_3$=75.3 atm cm⁹ gm⁻³, $\Pi$ is the osmotic pressure in Atmospheres, $m_W = 6.7 \times 10^5$ g/mol is the dextran molecular weight in our experiments, and $m_{W0} = 3.7 \times 10^5$ g/mol the dextran molecular weight used for the published measurement[41]. The term $m_{W0}/m_W$ corrects the van't Hoff coefficient for the molecular weight difference. Note that the osmotic pressure is relatively insensitive to molecular weight at experimentally relevant concentrations[41-43]. Importantly, this relationship shows that the osmotic pressure is non-ideal, exceeding the van't Hoff ideal gas limit by more than an order of magnitude.

The above described model assumes in plane hexagonal order. This is strictly applicable only in the crystalline phase, although there will still be local hexagonal order in the liquid phase. Plotting the experimental data for *fd-wt* and *fd*-Y12M against the theoretical prediction for the applied pressure reveals qualitative agreement in several respects (**Fig. 11**). First, the theory captures the apparent exponential decay, with decay constant $\kappa_{\text{eff}}^{-1} \approx 1.07$ nm, in agreement with the experimental decay length, $\kappa_{\text{eff}}^{-1} \approx 1.04$ nm (**Fig. 12**). These are only apparent decay lengths since the local Debye length is a function of the inter-rod spacing. Measuring pressure rather than force incurs an extra dependence on inter-rod spacing. Applying the same analysis to the force, $f = 3^{1/2} d_{\text{rod}} \Pi_{\text{int}}(d_{\text{rod}})$, yields apparent decay lengths of 1.16 nm and 1.20 nm for experiment and theory, respectively. Second, the theory reproduces the observation that the measured equation of state is nearly insensitive to persistence length (for $l_p \geq L$) in the limit of small inter-rod spacing, or equivalently high osmotic pressures. For comparison, the theoretical result in the rigid rod limit is also shown. In this regime, filament undulations are suppressed to scales that are smaller than the Debye length.

The theory over-estimates the osmotic pressure even in the rigid rod limit by a factor of ~5. This can be attributed to the the use of the Poisson-Boltzmann model that neglects the ion correlations, or neglecting the nonuniformity of the fixed charges in the viruses, or to the approximate calculation of the effective linear charge density. Since the interactions are quadratic in the effective charge, the discrepancy between theory and experiments would be eliminated by reducing the effective charge to about half its value. We performed an alternative calculation in which the viruses are treated as semiflexible filaments with hard-core interactions, where the effect of the electrostatics is accounted for by an "effective diameter" that is larger than the intrinsic virus diameter. However, this approach failed to capture the scaling of the experimental data with $d_{\text{rod}}$ or the lack of dependence on persistence length.

It is possible to estimate the location of the melting transition for an array of semiflexible chains interacting through hard-core interactions[44]. Extending this model to include electrostatic interactions is challenging and beyond the scope of the present work[32]. However, we can make a qualitative comparison to the Lindemann criteria for a 3D crystal of point particles, which states that melting occurs when the typical size of particle fluctuations reaches 0.1 $a$, with $a$ the lattice spacing [44]. The onset of the melting transition and its dependence on persistence length can be roughly captured by equating the scale of undulations (Eq. 4) to the inter-surface separation between rods according to $0.17(d_{\text{rod}} - \sigma)$, where the factor 0.17 was fit by eye (**Fig. 13**). However, this comparison is qualitative at best, and the melting transition predicted previously required undulations of about five times the Lindemann criteria[44].

**Discussion:** Osmotic stress techniques have been used to interrogate properties of diverse soft materials, providing insight into the molecular interactions that govern behaviours of both lipid bilayers and biopolymer suspensions[45-47]. Our method has a few differences with the standard implementation of the osmotic stress techniques. First, instead of extracting the average filament separation from X-ray scattering, we obtained the rod density by measuring the assemblage area with optical microscopy. Since there is no significant exchange of rods on experimental times scales, the membrane area is directly related to the filament spacing. Second, our technique permits *in-situ* change of the osmotic stress, which yields more accurate measurements. This feature revealed a discontinuity in the equation of state that is associated with a first other phase transition from a liquid to a solid state. In a similar spirit, a recent study has utilized the temperature dependence of poly(ethylene glycol) to extract the discontinuous change in the filament spacing associated with the transition from the cholesteric to line hexatic[48]. Finally, the ability to tune the osmotic pressure *in situ* and compatibility with optical microscopy enabled visualization of the kinetic pathways by which the metastable 2D colloidal structures transform into 3D materials.

Equilibrium colloidal membranes form over a fairly limited range of depletant concentrations. Osmotic pressures beyond an upper critical value leads to the formation of bulk smectic phases, while pressures below a lower critical value lead to nematic tactoids[49, 50]. Our microfluidic technique allowed for investigations of colloidal membranes over a wider range of osmotic pressures. These experiments demonstrated the fundamental difference between colloidal membranes and conventional 3D materials. Formation of either 3D tactoids or bulk smectic phases from a 2D colloidal membranes requires a rearrangement in which rods collectively escape into the third dimension. Such transformations have large nucleation barriers; consequently, colloidal membranes are highly metastable over a wide range of osmotic pressures. With decreasing pressure, a membrane can either directly evaporate into an isotropic phase or melt into a 3D nematic tactoids. Tactoid nucleation always takes place at the membrane's edge. This can be rationalized as the boundary conditions enforce uniform twist of the membrane's edge that locally melts a smectic monolayer into a nematic[1, 24]. Similar pathways were observed in mixtures viruses and thermosensitive polymer[51, 52].

*Intermolecular interactions:* Confined liquid crystalline semi-flexible filaments have more restricted degrees of freedom than filaments in a disordered isotropic suspension. Consequently, there is an entropic penalty associated with the formation of orientationally ordered phases. This penalty is smaller for more rigid filaments; thus, the stiffer *fd*-Y21M filaments form membranes at lower osmotic stress, when compared to the more flexible *fd-wt*. Our results suggest that theories of colloidal membranes phase behaviour which do not account for semi-flexibility will under predict the osmotic pressure required to stabilize membranes. This is consistent with previous findings that increasing filament flexibility supressed the formation of both nematic and smectic liquid crystals[20, 53-58]. Intriguingly, the filament flexibility does not influence the equation of state at higher osmotic pressures or equivalently smaller filaments separations. Recent experiments using classical osmotic stress techniques examined the interactions of both *fd-wt* and *fd*-Y21M virus, reached the same conclusion[40]. While our experiments studied a different range of osmotic stresses, our data is consistent with these results.

A Poisson-Boltzmann calculation that accounts for the increased density of counter ions within the membrane (compared to bulk) explains some features of the measured equation of state in the crystalline regime. This result indicates that electrostatics are the dominant interaction between rods at surface-surface separation distances in the range 3-6 nm. Consistent with the experimental results, the theory finds that filament conformational fluctuations have a small effect on rod interactions at inter-rod separations in the crystalline regime. We found that theories which represent electrostatic effects through hard-core interactions and an "effective diameter" could not describe the experimental measurements. Consistent with this observation, previous models of colloidal membranes and rafts that did achieve quantitative agreement with experiments without accounting for electrostatics assumed that the depletant osmotic pressure is given by an ideal equation of state[5, 59]. In fact, the empirically measured equation of state for dextran (Eq. 9) exceeds the ideal formula by an order of magnitude at experimentally relevant concentrations.

*Estimating elasticity of colloidal membranes:* The measured equation of state provides insight into the energetic cost of elastically deforming colloidal membranes. Similar to lipid bilayers, the out-of-plane deformations of colloidal membranes are described by the Helfrich energy that contains two parameters, Gaussian and mean curvature moduli[60]. The Gaussian modulus of colloidal membranes, $\bar{\kappa}$, has been measured using two independent methods[61, 62]. Both methods yielded $\bar{\kappa}$ that is positive and of the order of ~100 $k_BT$, suggesting that membranes can lower their free energy by adopting saddle-splay configurations. An early study of colloidal membranes estimated that $\kappa$ is also a few hundred $k_BT$[2]. The equation of state yields an independent estimate of the curvature modulus, by calculating the lateral membrane compressibility, $k_a \equiv -\frac{1}{A}\frac{\partial A}{\partial P}$. The compressibility is related to the bending modulus, $\kappa$, by $\kappa = k_a h^2/12$, where *h* is the membrane thickness. The equation of state measurements yield that $k_a \sim 230{,}000\ k_BT/\mu m^2$ at 40 mg/ml dextran concentration, which leads to a mean curvature modulus of $\kappa \sim 15{,}000\ k_BT$, where we assumed that membrane thickness is ~0.9 $\mu$m.

The equation of state based estimate of $\kappa$ is very different from the previous measurement[2]. In the original experiments, $\kappa$ was estimated by analysing the out-of-plane height fluctuations of membranes that were viewed in the edge-on configuration. However, the distance from the membrane edge was not systematically controlled and the height fluctuations were visualized by non-confocal techniques that do not effectively eliminate out-of-plane signal contributions. It is possible that the out-of-plane fluctuations studied previously were actually ssociated with the soft edge modes rather than the intrinsic bending modes. Recently, it has been shown that colloidal membranes exhibit significant out-of-plane edge-bound modes that create saddle-splay deformations. These fluctuations are driven by the low and positive value of the Gaussian modulus of colloidal membranes, and decay as one moves away from the edge[62]. Both the edge fluctuations and a 1D cross-section of the height fluctuations scale as ~$1/q^3$. The only way to rigorously distinguish between the two fluctuation modes is to analyse how the measured spectrum changes as one moves away from a free edge.

The conflicting measurements of curvature modulus demonstrate that the continuum properties of colloidal membranes are not well understood and further experimentation is required. In particular, there is a need for methods that use external force to robustly perturb a colloidal membrane structure in order to measure its bending rigidity. Alternatively, re-examining the height fluctuations of a colloidal membrane viewed in the edge-on configuration, and how they depend on the distance away from the edge, might provide additional insight into the microscopic origin of previously measured bending fluctuations.

**Acknowledgements:** We thank Tom Powers for useful discussions, and Achini Opathalage for suggestions related to the design of the microfluidic device. We acknowledge support of National Science Foundation through grants (NSF-MRSEC-1420382 and NSF-DMR-1609742). We also acknowledge use of Brandeis MRSEC biological synthesis facility, microfluidics fabrication facility and optical microscopy facilities supported by NSF-MRSEC-1420382. TG thanks ID02 staff at the ESRF for their help with SAXS experiments.

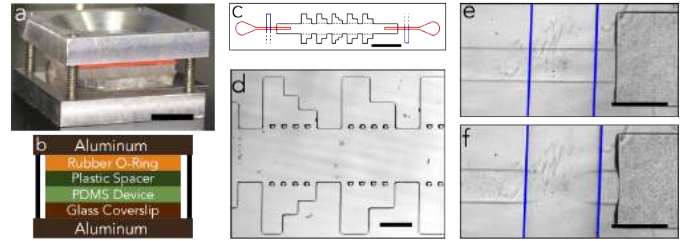

**Figure 1 | The microfluidic device for buffer exchange (a)** Image of a fully sealed and assembled microfluidic device. The PDMS chip is clamped to a glass coverslip, using two pieces of aluminium and four screws, to ensure no leakage between the two layers. **(b)** A schematic cross section of the microfluidic device with individually labelled layers. A rubber O-ring and plastic spacer distribute the clamping pressure across the PDMS chip evenly. **(c)** Flow is from right to left. A top-down view of the device design debossed on the PDMS chip. The design has an inlet and outlet (red), a flow channel with storage notches (black), and two Quake valves (blue) that control the flow through the channel. Small circular posts, placed at the entrance of each notch, combined with a step-shaped design, reduce laminar flow velocity in the notches. **(d)** Flow is from right to left. A brightfield image of a PDMS device filled with depletant and buffer. The field of view is at the centre of the flow channel. **(e,f)** The valve that controls the flow within the channel, shown in open and closed configuration. Scale bars, 1 cm (a), 1.5 mm (c), and 250 μm (d,e,f).

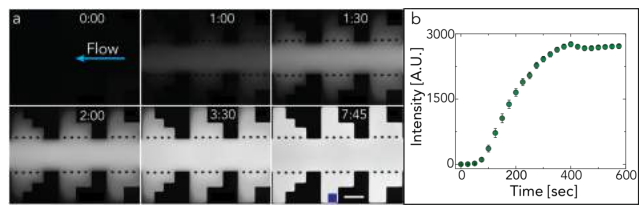

**Figure 2 | Time delay associated with buffer exchange (a)** The microfluidic exchange device is viewed with fluorescence microscopy. The fluorescent depletant and buffer mixture is flowed in from the right side of the page. Both valves are in the open configuration. The time after flow was started is presented on each frame as *min:sec*. Measurements are performed in a small area of a one-step notch (shaded in blue in the last time frame). **(b)** Pixel intensity measures the concentration of the new mixture in the device, relative to the original sample concentration. The complete buffer exchange occurs after the intensity plateaus. The mean pixel intensity, with background subtraction, was measured in the shaded blue region as a function of time. Scale bar, 250 μm.

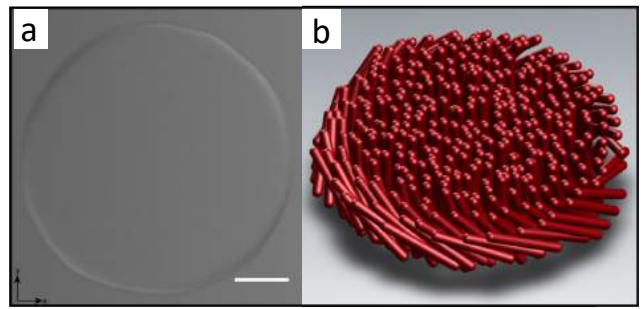

**Figure 3 | Three views of colloidal membranes**. **(a)** A top-down view of a 2D colloidal membrane imaged using DIC microscopy. **(b)** A schematic of a colloidal membrane. There is complete phase separation between the self-assembled rod-shaped viruses (red) and depleting polymer, which maximizes the system entropy (polymer is not shown). Scale bar, 5 μm.

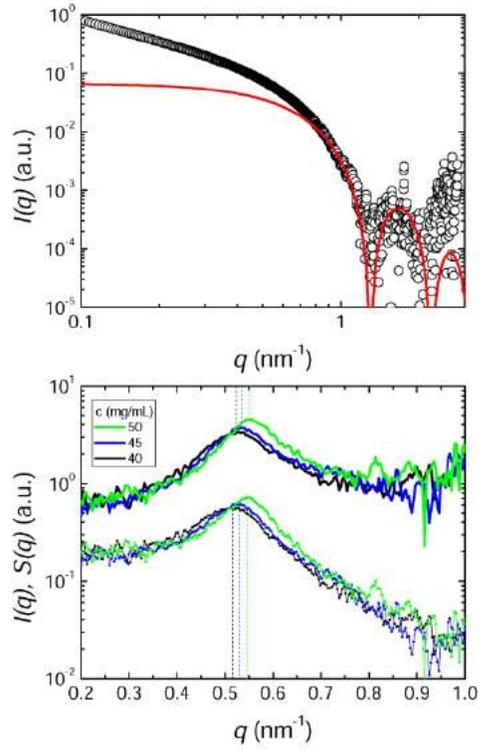

**Figure 4 | Small Angle X-Ray Scattering on Colloidal Membranes.** (a) Form factor of an isotropic dispersion of fd-virus at 0.5 mg/mL (circle). Corresponding theoretical form factor $F$ of the fd-virus all aligned along the x-ray beam. (b) Intensity scattered by a colloidal membrane (bottom curves) and structure factor $S=I/F$ (top curves) with $F$ being the theoretical form factor. The position of the peak is indicated by the vertical dash line. Going from $I$ to $S$ and increasing dextran concentration $c$ shift $q_{peak}$ from 0.517 to 0.530, 0.530 to 0.540 and 0.545 to 0.553 nm$^{-1}$.

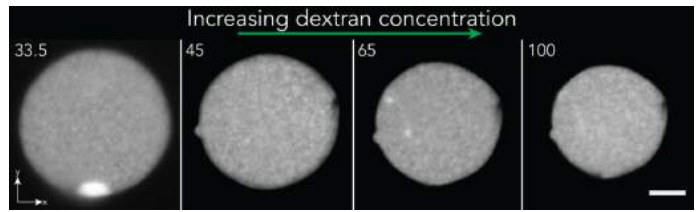

**Figure 5 | Membrane area decreases with increasing osmotic pressure.** A series of images of a particular membrane, composed of fluorescently labelled $fd$-Y21M rods, shown at four different dextran concentrations. Numerical labels indicate dextran concentration in mg mL$^{-1}$. Scale bar, 5 μm.

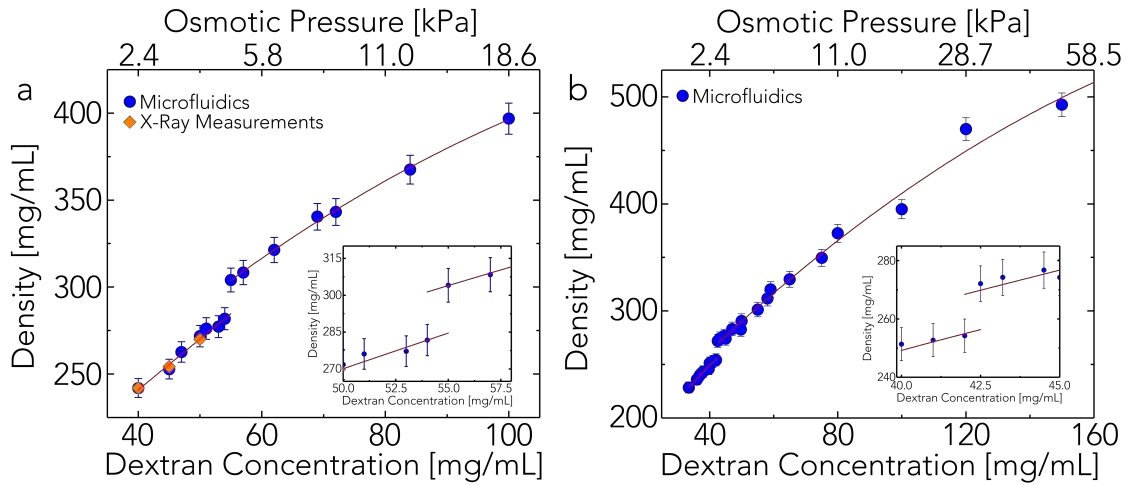

**Figure 6 | Equation of state of fd-wt and fd-Y21M colloidal membranes.** (a) Rod density as a function of dextran concentration measured for *fd-wt* membranes. Inset: Discontinuity in the equation of state at ~54 mg/ml is consistent with a first order freezing and phase transition between a 2D liquid-like and 2D solid-like membrane. Diamonds indicate three samples where the in-plane rod density was determined with small angle X-ray scattering measurement. Red bars indicate regions where colloidal membranes are equilibrium structures. (b) Equation of state for fd-Y21M membranes. The membrane freezes at smaller osmotic stresses. The orange lines are guides to the eye. Error bars are the standard deviation of 10 measurements of membrane area.

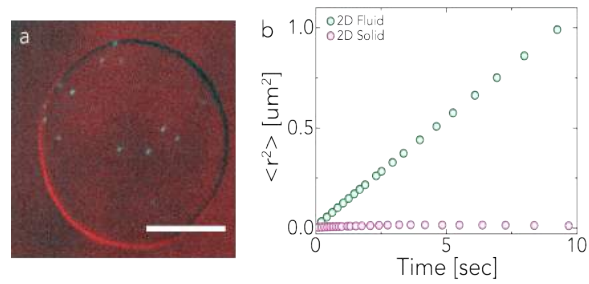

**Figure 7 | Dynamics of viruses in liquid and solid membranes** (a) A top-down view of a *fd-wt* membrane (red) taken with DIC microscopy is overlaid with a fluorescence image that shows isolated viruses (cyan). This technique was used to visualize individual rods moving in the bulk (Supplementary Movie #). (b) The mean square displacement of fluorescently labelled rods was measured as a function of time for both a 2D liquid-like (green) and a 2D solid-like (purple) membrane. Rods in the liquid-like membrane exhibit diffusive behaviour with a diffusion coefficient of .03 μm/sec, while those in the frozen membrane have no measurable motion. Scale bar, 5 μm.

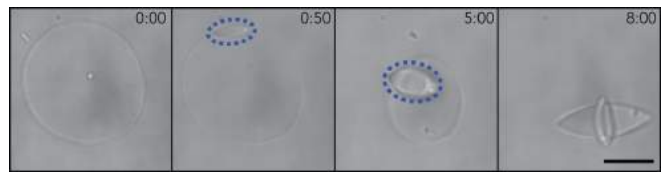

**Figure 8 | 2D colloidal membrane melts into a 3D nematic tactoids.**
A time lapse showing how a metastable 2D colloidal membrane melts into a stable 3D tactoid (blue dashed outline) at low osmotic pressure. The tactoid nucleates from the membrane's twisted edge, where nematic order is already very high. The tactoids exhibit a surface frozen smectic monolayer. Scale bar, 10 μm.

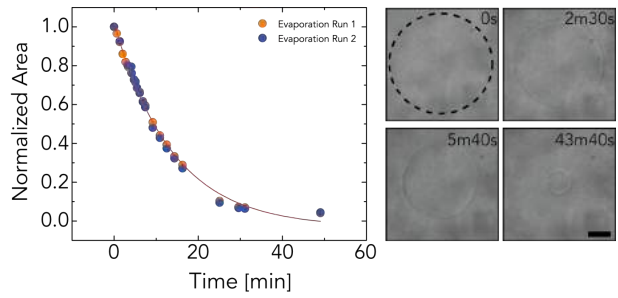

**Figure 9 | Evaporation of colloidal membranes at low osmotic pressures. (a)** Membrane area as a function of time was measured for a variety of different membranes. Two curves were normalized such that evaporation begins at area equal to one, and experiments were performed at 25 mg mL$^{-1}$ dextran concentration. Each curve represents a unique membrane. All curves are fit well by exponential decay functions and exhibit an average of rate off $k_{off}$ rate of 4.1 min$^{-1}$. **(b)** A sequence of images showing how the membrane area shrinks with time due to evaporation of virus particles. Scale bar, 5 µm

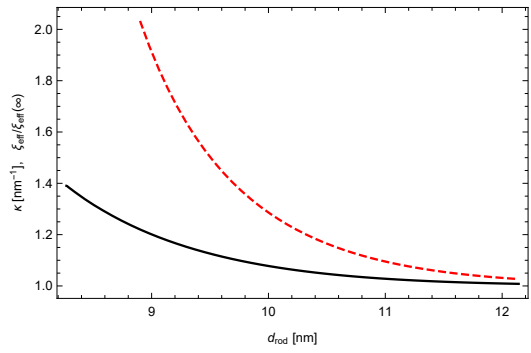

**Figure 10 | The electrostatic screening length and effective filament charge within colloidal membranes.** The inverse Debye length, $\kappa$ (black line), and the corresponding dimensionless effective linear charge density $\xi_{eff}$ (dashed red line) plotted as a function of inter-rod separation. The effective charge density is normalized by its bulk value, $\xi_{eff} = 36.15$ for $\kappa_D = 1/nm$.

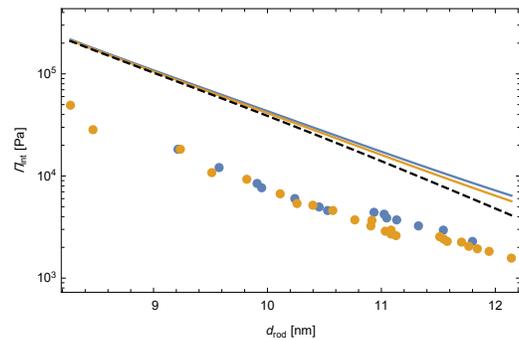

**Figure 11 |** The theoretical predictions for the osmotic pressure as a function of inter-rod separation are shown for persistence lengths corresponding to *fd*-wt ($l_p = 2$ µm, blue line) and *fd*-Y121M ($l_p = 10$ µm, orange line), overlaid on the experimental data. The theoretical results are obtained from Eqs. 4-9. The dashed black line shows the result in the rigid rod limit ($l_p = \infty$).

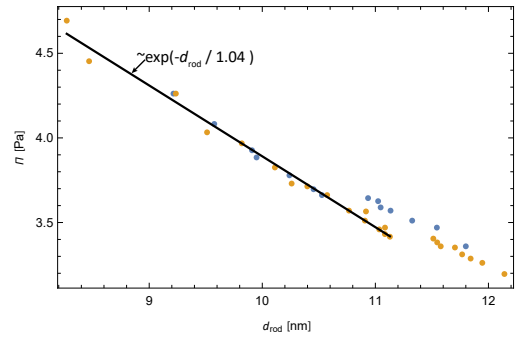

**Figure 12 | Osmotic pressure of crystalline membranes scales exponentially with filament separation.** The experimentally measured equation of state for *fd-wt* and *fd*-Y121M is plotted as a function of the inter-rod separation. The log scale *Y*-axis illustrates the exponential decay. The solid black line is a best fit to the data within the crystalline regime, yielding a decay length of $\kappa_{\text{eff}}^{-1} = 0.85$ nm$^{-1}$.

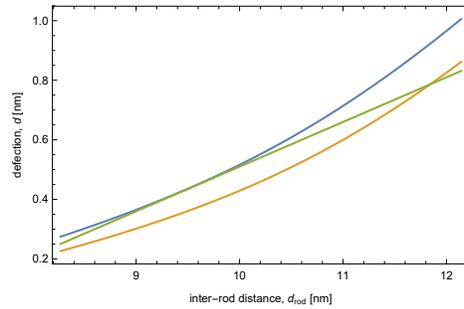

**Figure 13 |** The scale of filament undulations, $d$, calculated from Eq. 4 is shown for *fd-wt* (blue line) and *fd*-Y121M (orange line) as a function of inter-rod separation. The straight line is an effective Lindemann criteria, $d = 0.17(d_{\text{rod}} - \sigma)$, which estimates the inter-rod separation at which the membrane melts for each case.